\documentclass[preprint,showkeys,aps,prl]{revtex4}
\usepackage[dvips]{graphicx}
\begin{document}

\title{Hamiltonian Thermostats Fail to Promote Heat Flow}

\author{
Wm. G. Hoover and Carol G. Hoover               \\
Ruby Valley Research Institute                  \\
Highway Contract 60, Box 601                    \\
Ruby Valley, Nevada 89833                       \\
}

\date{\today}


\vspace{0.1cm}

\begin{abstract}
Hamiltonian mechanics can be used to constrain {\it temperature} simultaneously with  energy.  We illustrate the interesting situations that develop when two {\it different} temperatures are imposed within a composite Hamiltonian system. The model systems we treat are $\phi^4$ chains, with quartic tethers and quadratic nearest-neighbor Hooke's-law interactions.  This model is known to satisfy Fourier's law.  Our prototypical problem sandwiches a Newtonian subsystem between hot and cold Hamiltonian reservoir regions.  We have characterized four different Hamiltonian reservoir types.  There is no tendency for any of these two-temperature Hamiltonian simulations to transfer heat from the hot to the cold degrees of freedom.  Evidently steady heat flow simulations {\it require} energy sources and sinks, and are therefore incompatible with Hamiltonian mechanics. 
\end{abstract}

\maketitle

\section{Introduction}

Molecular dynamics has facilitated the modeling of the macroscopic equation of state and constitutive relations in terms of atomistic Hamiltonians.  Three research groups developed the method\cite{b1,b2,b3} and its applications:  Fermi's one- and two-dimensional simulations at Los Alamos, Vineyard's radiation-damage simulations at Brookhaven, and Alder's $H$-Theorem and phase-change studies at Livermore.  All three groups treated the approach to equilibrium from nonequilibrium initial conditions.  Such problems are straightforward applications of Newtonian, or Lagrangian, or Hamiltonian mechanics.  Today both the scale and the complexity of the modeling have expanded to make molecular dynamics a comprehensive tool for learning and understanding.

Some conceptual problems remain\cite{b4}.  Irreversibility, and its connection to Lyapunov instability -- the exponential growth of small perturbations -- are undergoing wide-ranging investigations.  Nonequilibrium boundary conditions have been under intensive development\cite{b5,b6}.  From a theoretical standpoint there is no agreement on the proper definition of nonequilibrium states.  How should they be described?  Nonequilibrium {\it temperature}\cite{b7,b8,b9} can appear to be a difficulty.  The many equivalent equilibrium definitions, thermodynamic, kinetic, and configurational, all differ from one another away from equilibrium, with the latter two temperatures becoming tensors rather than scalar quantities\cite{b10}.  In strong shockwaves the longitudinal and transverse kinetic temperatures can differ temporarily by an order of magnitude\cite{b11}, but then equilibrate in a few collision times.  Sometimes the local configurational temperature is {\it negative}\cite{b12}.

{\it Stationary} nonequilibrium flows require nonequilibrium boundary conditions to impose local velocities and temperatures and to extract the irreversible heat generated by equilibration processes\cite{b5,b6}.  The Nos\'e-Hoover version of Nos\'e's constant-temperature dynamics provides a robust time-reversible approach to temperature control.  That approach uses integral feedback.  Isokinetic velocity rescaling\cite{b13} can likewise be described with time-reversible deterministic motion equations using differential feedback.  Couette shear flow, Fourier heat conduction, and shockwave propagation are prototypical examples of nonequilibrium problems which can all be driven and maintained by special boundary regions with stationary velocities and temperatures\cite{b4}.

Because there are many definitions of temperature it is natural to explore their relative usefulness in driving flows away from equilibrium.  Both kinetic and configurational temperatures can be controlled by constraining Hamiltonian systems.  Here we compare five such approaches to the simplest heat-flow problem, conduction in a $\phi^4$ chain\cite{,b14,b15,b16,b17}.  Such a chain combines harmonic Hooke's-Law interactions with quartic tethering potentials.  The $\phi^4$ chain is an improved relative of the seminal anharmonic chain models studied by Fermi's group at Los Alamos.  His models lacked the realism of the tethered-particle simulations studied in much greater detail fifty years later, by Aoki and Kusnezov.  Just as Fermi's results (lack of equilibration) were a surprise to Fermi, the present results (once again, lack of equilibration!) surprised {\it us}.  What we find is apparently a common and serious drawback of Hamiltonian thermostats, a failure to promote heat flow\cite{b17}.  The failure of recent workers to recognize this defect\cite{b18} helped motivate the present work.

The body of this paper has three parts.  In Section II we describe the thermostats to be considered.  In Section III we compare the results of sandwich simulations of [ cold + Newtonian + hot ] thermostated systems.  Section IV is a summary of our findings and the conclusions which we draw from them.

\section{Several Hamiltonian Thermostats}

Nos\'e\cite{b19,b20,b21} took a bold step forward in 1983, finding a Hamiltonian mechanics which could model the canonical-ensemble isothermal distribution.  His Hamiltonian, for $\#$ degrees of freedom , among which is a new ``time-scaling'' variable $s$ (we use $s^2$ here to ensure that the logarithm is meaningful) along with its conjugate momentum $p_s$ , is :
$$
{\cal H}_{\rm Nos\acute{e}} = \sum (p^2/2ms^2) + \Phi(q) + (p_s^2/2M) + (\# kT/2)\ln(s^2) \ .
$$
The adjustable parameter $M$ can be used to vary the timescale of the $( \ s,p_s \ )$ thermostat.  Nos\'e showed that the dynamics for this $( \ q,p,s,p_s \ )$ Hamiltonian was consistent with the stationary Gibbsian canonical distribution for the coordinates and the ``scaled momenta'' $\{ \ q,(p/s) \ \}$ :
$$
{\cal H}_{\rm Nos\acute{e}} \longrightarrow f(q,p,s) \propto \exp[ \ -(K(p/s)/kT) - (\Phi(q)/kT) \ ] \ .
$$
A simpler route\cite{b22} to a similar result is to begin with the Nos\'e-Hoover motion equations, a modification of Nos\'e's work in which $s$ is completely absent and $p_s$ is replaced by a time-reversible friction coefficient, $\zeta $, along with its characteristic relaxation time $\tau$ :
$$
\{ \ \dot q = (p/m) \ ; \ \dot p = F(q) - \zeta p \ \} \ ; \ \dot \zeta = \sum [ \ (p^2/mkT) - 1 \ ]/\tau^2 \ .
$$
Next, it is easy to see that a {\it Gaussian} distribution for $\zeta $, along with the canonical distribution for the $( \ q,p \ )$ variables satisfies the phase-space continuity equation (a generalized Liouville Theorem), giving the result :
$$
(\partial f/\partial t) = 0 \longrightarrow
 f(q,p,\zeta) \propto \exp[ \ -(K(p)/kT) - (\Phi(q)/kT) -(1/2)(\zeta \tau)^2\ ] \ .
$$
$\zeta $ controls the flow of energy to and from the thermostated system with the arbitrary relaxation time $\tau $.  Bauer, Berry, Bra\'nka, Bulgac, Hamilton, Jellinek, Klein, Kusnezov, Martyna, Tuckerman, Winkler, and Wojciechowski all suggested various modifications of these motion equations\cite{b23,b24,b25,b26,b27,b28,b29}.  Surprisingly, Bulgac and Kusnezov were able to demonstrate the feasibility of modeling Brownian Motion with strictly {\it time-reversible} deterministic motion equations by introducing three thermostat variables like $\zeta$ , rather than just one or two\cite{b30}.

Lagrangian and Hamiltonian mechanics suggest a wide variety of approaches to thermostating.  Dettmann and Morriss\cite{b31,b32} discovered Hamiltonian bases for both the Nos\'e-Hoover and Gaussian isokinetic motion equations.  Bond, Laird, and Leimkuhler rediscovered this approach a year later\cite{b33}.  [ The isokinetic equations are the instantaneous $\tau \rightarrow 0$ limit of the Nos\'e-Hoover motion equations given above. ]  Landau and Lifshitz' ``configurational temperature''\cite{b34} was rediscovered and generalized by Braga, Rugh, and Travis\cite{b7,b8,b35} ,
$$
kT_c \equiv \langle F^2 \ \rangle/\langle \ \nabla^2{\cal H} \ \rangle \ .
$$
Constant configurational temperature can be imposed as a straightforward ( and tedious ) holonomic constraint\cite{b7,b8,b17}.

Recently Campisi, H\"anggi, Talkner, and Zhan suggested the use of a logarithmic thermostat\cite{b36}, very much like Nos\'e's, but without an explicit coupling to the remaining degrees of freedom :
$$
{\cal H}_{\rm CZTH} = (p_s^2/2m) + (kT/2)\ln(s^2 + \delta^2) \ . 
$$
Here we include a $\delta^2$ in the definition so as to avoid divergence when $s$ changes sign. The equations of motion of the logarithmic thermostat, absent the coupling forces linking it to the system it influences, are :
$$
\dot s = (p_s/m) \ ; \ \dot p_s = -skT/(s^2 + \delta^2) \longrightarrow
$$
$$
\langle \  s\dot p_s \ \rangle = \langle \ (d/dt)(sp_s) \ \rangle  - \langle \ \dot sp_s \ \rangle =
-\langle \ \dot sp_s \ \rangle \simeq -kT \ .
$$
Notice that the time-averaged derivative of a bounded quantity, here $(sp_s)$, vanishes.  This averaging operation shows that when $\delta$ can be ignored the time-averaged kinetic energy of the ``thermostat'' is $\langle \ (p_s^2/2m) \ \rangle = (kT/2)$ .  The relatively poor results obtained with this thermostat were pointed out by Espa\~nol, Hoover, and Mel\'endez and documented in a series of ar$\chi$iv contributions\cite{b18,b37,b38}.  It is worth noting here that the instantaneous configurational temperature of the pure logarithmic-thermostat potential is negative, $-kT$ !
$$
\phi(s) = kT\ln(s) \rightarrow \{ \ F(s) = (-kT/s) \ ; \ \nabla^2 \phi = (-kT/s^2) \ \} \rightarrow kT_c \equiv -kT \ !
$$
This bipolar character, with $T_c = -\langle \ T_k \ \rangle $, is definitely uninviting.

  An isokinetic form of mechanics, based on a nonholonomic ( velocity-dependent ) constraint, can be imposed by using the Hoover-Leete Lagrangian and Hamiltonian\cite{b17,b39},
$$
{\cal L}_{\rm HL}( \ q,\dot q \ ) \ \rightarrow \ K(\dot q) \equiv
\sum (m\dot q^2/2) {\rm \ constant \ } \longrightarrow
$$
$$
{\cal H}_{\rm HL}( \ q,p \ ) \equiv 2\sqrt{K(p)K(\dot q)} - K(\dot q) + \Phi(q) \ .
$$
This Lagrangian approach closely resembles the Gauss'-Principle isokinetic approach, though the resulting trajectories are quite different.  In the Hoover-Leete approach the momenta $\{ \ p \ \}$ evolve in the usual way, but the velocities are continuously rescaled, by a Lagrange multiplier $\lambda$, so that there are two different versions of the kinetic energy :
$$
\{ \ \dot q = (p/m)/(1+\lambda) \ ; \ \dot p = F \ \} \ ;
\ K(p) \equiv \sum (p^2/2m) \neq  \sum (m\dot q^2/2) \equiv K(\dot q)\ .
$$   

In the remainder of this work we apply the original Nos\'e-Hoover thermostat, Nos\'e's thermostat, the logarithmic thermostat, and the Hoover-Leete thermostat, to a simple model system known to follow Fourier's law, the $\phi^4$ model, so called because each particle in a harmonic chain is, in addition, tethered to its lattice site by a quartic potential.  Aoki and Kusnezov carefully characterized the $\phi^4$ model's dependence on temperature\cite{b14,b15}.  With the masses and both force constants and Boltzmann's constant all set equal to unity the $\phi^4$ model has a heat conductivity $\kappa \simeq 2.8/T^{4/3}$ in one dimension\cite{b15}.

\section{Comparing Five Conductivity Approaches in $\phi^4$ Chains}

The $\phi^4$ model can be implemented in any number of dimensions and with any lattice structure.  Likewise the number of particles and the size of the temperature gradient can be large or small.  Fortunately Aoki and Kusnezov have carried out a series of comprehensive investigations establishing that the model {\it is} a useful representation of Fourier conductivity over a wide range of conditions and dimensionality.  The temperature profiles they found have a typical ``jump'' at system boundaries but are otherwise quite unremarkable\cite{b14,b15}.  For simplicity we confine all of our investigations reported here to a 60 particle chain, with twenty Newtonian particles sandwiched between a cold 20-particle reservoir, at temperature 0.5 and a hot 20-particle reservoir at a temperature 1.5.  We choose an overall energy such that the Newtonian particles start out with an average temperature of order 1.0.  Except for the Nos\'e-Hoover case, all of the simulations carried out here have constant total energy, due to their Hamiltonian character.

\subsection{Nos\'e-Hoover Thermostat}

\begin{figure}
\includegraphics[height=11cm,width=8.5cm,angle= -90]{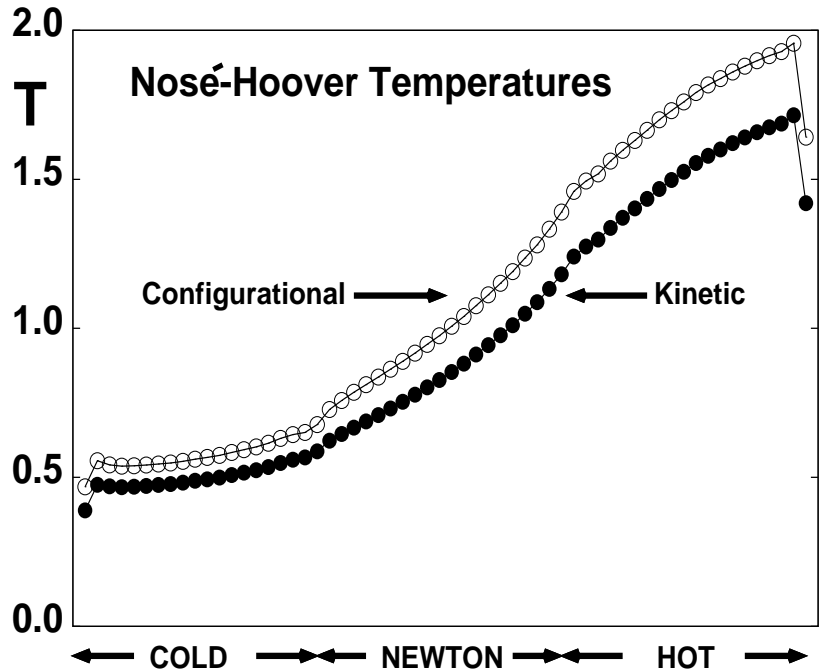}
\caption{
Kinetic and configurational temperature profiles for a 60-particle $\phi^4$ chain according to the Nos\'e-Hoover equations of motion.  The particle temperatures are averages over a billion-timestep simulation with $dt = 0.01$.  By ``billion'' we use the word to mean one thousand million, $10^9$, throughout this work.  Here, as in {\it all} of our simulations, we use the classic fourth-order Runge-Kutta integrator.  The time-averaged heat flux is -0.0618, with the flow from right to left.  This corresponds to a heat conductivity $\kappa \simeq 2$.  Simulations with longer chains are in agreement with Aoki and Kusnezov's work cited in References 14 and 15.
}
\end{figure}

The straightforward approach to this problem uses the ( {\it nonHamiltonian} ?) Nos\'e-Hoover equations of motion\cite{b21,b22} in both 20-particle reservoir regions :
$$
\{ \ \dot q = (p/m) \ ; \ \dot p = F(q) - \zeta p \ \} \ ;
\ \dot \zeta = [ \ \langle \ (p^2/mkT) \ \rangle - 1 \ ]/\tau^2  \ .
$$
In our numerical work the particles have unit mass.  Also, both Boltzmann's constant $k$ and the thermostat relaxation times $( \ \tau_{\rm cold},\tau_{\rm hot} \ )$ have been chosen equal to unity.  We indicate the questionable nonHamiltonian character here, in order to emphasize that Dettmann\cite{b31} (and later, Dettmann with Morriss\cite{b32}) found and described a special Hamiltonian which, when set equal to zero, generates these same equations of motion, but only for a single temperature --- {\it not} simultaneously for two or more of them.

At any time $t$ the average values of $p^2$ for the two reservoir regions, indicated here by $\langle \ p^2 \ \rangle$ , are one-twentieth the sum of the 20 instantaneous cold or hot reservoir-particle contributions.  The forces in this problem are of two kinds: nearest-neighbor Hooke's-Law forces and cubic forces from the quartic tethering potential :
$$
\Phi = \sum (1/2)(x_{i+1} - 1.0 - x_i)^2 + \sum (1/4)(x_i - x_{oi})^4 \ ,
$$
where the fixed sites $\{ \ x_o \ \}$ have a regular lattice spacing of unity.  Likewise the particle masses, force constants, relaxation times, and Boltzmann's constant have all been set equal to unity, for simplicity.

Figure 1 shows the kinetic and configurational temperature profiles that result from a billion-timestep fourth-order Runge-Kutta solution of the 20 ``cold'' plus 20 ``Newton'' plus 20 ``hot'' pairs of equations for $\{ \ \dot q,\dot p \ \}$ plus the additional two differential equations for the cold and hot friction coefficients $\{ \ \zeta_{\rm cold},\zeta_{\rm hot} \ \}$ .  The resulting kinetic temperature profile is unremarkable, save for the slight jumps at the system boundaries and thermostat interfaces.  The difference between the two temperature definitions, kinetic and configurational, is a measure of the need for clarity in defining nonequilibrium system properties.  Nos\'e-Hoover integral feedback guarantees precise temperature control.  The kinetic temperatures are {\it automatically} equal to the target temperatures ( 0.5 in the cold region -- 1.5 in the hot ).  This follows because the {\it long time averages} of the differential equations for the friction coefficients necessarily reproduce exactly the specified kinetic temperatures :
$$
\{ \langle \ \dot \zeta \ \rangle = 0 \longrightarrow \langle \ (p^2/mkT) \ \rangle \equiv 1 \ \} \ .
$$
Carlos Braga and Karl Travis developed a similar automatic feedback approach for configurational temperature\cite{b40}.
 The heat flux for this problem corresponds to a conductivity significantly smaller than the large system limit, $\kappa \simeq 2.8$, established by Aoki and Kusnezov.  We have confirmed that longer chains agree very well with Aoki and Kusnezov's work.  Let us use this 60-particle solution as a ``standard case'' for the $\phi^4$ model and investigate the same problem using four types of Hamiltonian heat reservoirs.

\subsection{Original Nos\'e Thermostat (with $s$ included in the Motion Equations)}
\begin{figure}
\includegraphics[width=4in,height=5in,angle= -90]{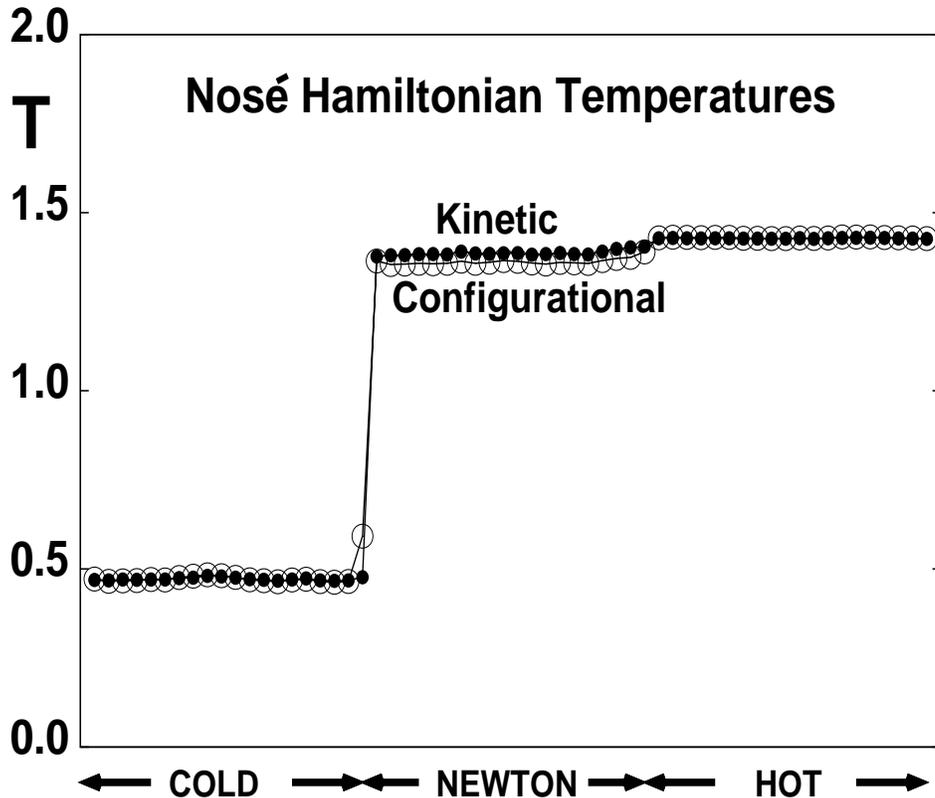}
\caption{
Kinetic and configurational temperature profiles using Nos\'e's original Hamiltonian {\it without time scaling} to constrain the temperatures of 20 ``Cold'' and 20 ``Hot'' particles.  In each of the three regions the configurational temperatures (open circles) are in good agreement with the kinetic temperatures.  This simulation fails to generate a heat flux from hot to cold.  The time averages come from a billion-timestep simulation with $dt = 0.0005$.  The value of the Hamiltonian is 15.000.
}
\end{figure}

In Nos\'e's revolutionary 1984 publications\cite{b19,b20} he introduced his original Hamiltonian, which contains an additional conjugate pair of Hamiltonian variables $(s,p_s)$ :
$$
{\cal H}_{\rm Nos\acute{e}} = \sum (p^2/2ms^2) + \Phi(q) + (\# kT/2)\ln(s^2) + (p_s^2/2M) \ .
$$
Again choosing the path of simplicity, we set the degrees-of-freedom parameter $\#$ and the thermostat's effective mass $M$ both equal to 20.  We set the particle mass $m$ and Boltzmann's constant both equal to unity.  The results which we find for Nos\'e's original Hamiltonian thermostats are shown in Figure 2.  The relative stiffness of the motion equations for the particles ,
$$
\{ \ \ \dot q = (p/ms^2) \ ; \ \dot p = F \ \} \ ; \ \dot s = (p_s/M) \ ;
\ \dot p_s = \sum^{\#} [ \ (p^2/ms^2)  - kT \ ]/s \ ,
$$
is a consequence of Nos\'e's time-scaling variable $s$ , which appears in the denominator.  The stiffness requires a much smaller timestep $dt$.  Problems with only a few degrees of freedom tend to be singular unless $(s^2)$ is replaced by $(s^2 + \delta^2)$, a precaution not needed here.  We used $dt = 0.0005$ in order to conserve energy to six-digit accuracy over the course of a billion-timestep run.  

Despite the extra work due to the smaller timestep $dt$, twenty times more for the same run duration, relative to the Nos\'e-Hoover equations, the temperature ``profile'' is disappointing.  It shows a lack of effective interaction between the thermostated regions and the Newtonian particles.  These results are {\it typical}\cite{b17}.  Higher or lower energies, or longer or shorter systems, provide similar results, establishing that Nos\'e's Hamiltonian is ineffective for heat flow problems.  The two ``time-scaling'' variables $( \ s_{\rm cold},s_{\rm hot} \ )$, with initial values of unity, had averaged values of 16 and 0.11 respectively, for the run shown in the Figure.  Although the temperature profiles, both kinetic and configurational, appear to be well converged, $s_{\rm cold}$ increases and $s_{\rm hot}$ decreases, though with significant fluctuations, throughout this relatively long run.  We turn next to the somewhat simpler appearing version of Nos\'e's idea promoted in Reference 36.

\subsection{``Weakly Coupled'' Logarithmic Thermostat}

\begin{figure}
\includegraphics[width=4in,height=4in,angle= -90]{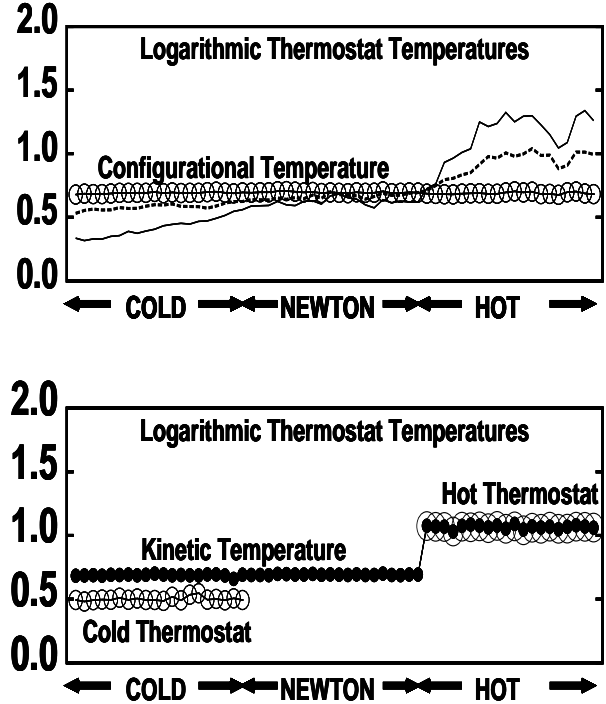}
\caption{
  The upper plot shows the configurational temperature profile averaged over times of 3000, 10000, and 100 000, with the last time shown as open circles.  Notice that the configurational temperature equilibrates well despite the discontinuities in the nearby constrained kinetic temperatures shown in the lower plot.  The 60 particle kinetic temperatures are represented by small filled circles.  The open circles in the lower figure are the kinetic temperatures of the Logarithmic-thermostat particles.  Notice that the ``hot'' thermostat particles have measured temperatures far below the specified $T=1.5$.  There is no measured heat flux in this simulation.  Two billion timesteps, with $dt = 0.00005$, were used to compute these averages.  The Hamiltonian is 52.882.
}
\end{figure}

About a year ago Campisi {\it et alii} [ CZTH ] introduced a logarithmic thermostat much like Nos\'e's but without an explicit coupling between the thermostat and the system\cite{b18,b36,b37,b38}.  So as to apply their idea to the sixty-particle $\phi^4$ chain we attach a CZTH logarithmic thermostat to {\it each} of the forty thermostated particles :
$$
{\cal H}_{\rm CZTH} = \sum^{60} (p^2/2m)_i + \sum^{40} [ \ (p_s^2/2M)_i + (kT_i/2)\ln \left((q_i-s_i)^2 + \delta^2\right) \ ] + \Phi(q) \ .
$$
The total number of differential motion equations to be solved is 200, four for each thermostated particle $( \ q,p,s,p_s \ )$ and two for each Newtonian particle $( \ q,p \ )$ .  In addition we use 100 more equations to compute the time integrals of  the 60 particle's kinetic temperatures and the 40 thermostat variables' kinetic temperatures.  The {\it configurational} temperatures ,
$$
kT_{\rm c} \equiv \langle \ F^2 \ \rangle / \langle \ \nabla^2 {\cal H} \ \rangle \ , 
$$
for the 60 particles require 120 additional equations, 60 for $\langle \ F^2 \ \rangle $ and another 60 for the particles' $\langle \ \nabla^2 {\cal H} \ \rangle $ .  Just as before we set Boltzmann's $k$ and all the particle masses equal to unity.  We also set the thermostat-variable masses $\{ \ M \  \}$ all equal to one.  We choose the ``small'' parameter $\delta = 0.01$ in order to avoid the singular behavior of purely-logarithmic thermostats.  Once again the logarithmic thermostat leads to stiff equations, requiring a timestep of 0.00005 for six-figure energy conservation in a two-billion-timestep run.

With the logarithmic ``thermostat'' we show separately both the long-time-averaged kinetic and configurational temperatures of the reservoir particles, $\langle \ p^2 \ \rangle$ and the kinetic temperatures of the corresponding thermostat variables $\langle \ p_s^2 \ \rangle $.   It is disconcerting to learn that the temperatures of the thermostat variables and the thermostated variables they are assigned to control can differ by a factor of two!  Otherwise the basic kinetic-temperature results are very like those found in our application of Nos\'e's original Hamiltonian work. All of the temperatures are shown in Figure 3.  There is no trace of a smooth temperature gradient like that generated by the Nos\'e-Hoover motion equations.  We conclude that the CZTH ``thermostat'' behaves much like Nos\'e's.  The logarithmic thermostats are evidently both ``stiff'' and ineffective.

\subsection{Hoover-Leete Isokinetic Thermostat}

\begin{figure}
\includegraphics[width=2.5in,angle= -90]{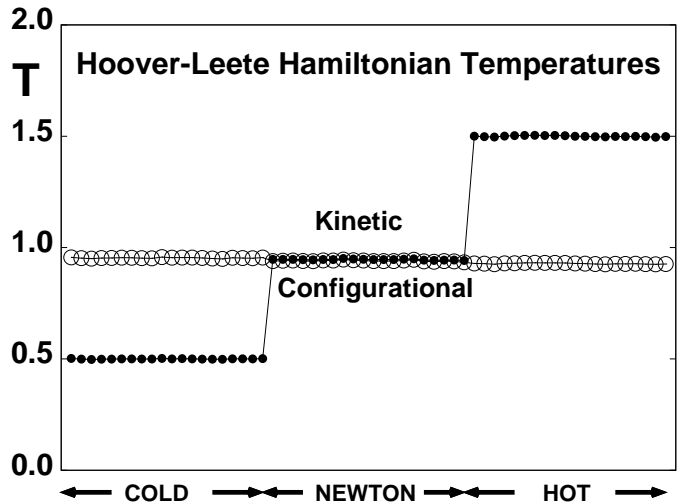}
\caption{
Time-averaged kinetic and configurational profiles using the Hoover-Leete Lagrangian/Hamiltonian thermostat to maintain the kinetic energies of the cold and hot regions.  Despite the successful thermostating of the reservoirs there is no heat flow.  The configurational temperatures, indicated by open circles, are nearly constant, and are equilibrated with the 20 Newtonian particles in the center of this $\phi^4$ chain.  This simulation includes a billion timesteps with $dt = 0.001$.  The value of the Hamiltonian is 49.282.
}
\end{figure}

One of us, Bill, met with Tom Leete where Tom worked, at the Aberdeen Proving Ground in Maryland.  Tom had chosen to work for the Army there because it was ``easier'' than graduate school.  This was shortly after Leete had finished his 1979 Master's Thesis work at West Virginia University, ``The Hamiltonian Dynamics of Constrained Lagrangian Systems''.  Bill was still in the throes of developing the thermostated nonequilibrium shear and heat flow treatments he had worked out with his first Ph. D. student William T. Ashurst.  Ashurst's thesis, ``Dense Fluid Shear Viscosity and Thermal Conductivity {\it via} Nonequilibrium Molecular Dynamics'', was published in 1974.  Bill worked out, with Tom's help, a Hamiltonian thermostat which included a nonholonomic constraint on the particle-based kinetic energy, $K(\dot q) = \sum(m\dot q^2/2)$ \ :
$$
{\cal H}_{\rm HL}( \ q,p \ ) = 2\sqrt{K(p)K(\dot q)} - K(\dot q) + \Phi(q) \ ; \ K(\dot q) \ {\rm constant} \ .
$$
The resulting Hamiltonian equations of motion allow the ``momenta'' $\{ \ p \ \}$ to change with time in the usual way, but constrain the summed up squares of the velocities $\{ \ \dot q \ \}$ to remain constant,
$$
\sum m\dot q^2 = 20kT \ ; \ \{ \ \dot q = (p/m)\sqrt{K(\dot q)/K(p)} \ ; \ \dot p = F(q) \ \} \ .
$$
Generally implementations of nonholonomic constraints lack uniqueness.  But a unique implementation results if one applies Gauss' Principle (of Least Constraint)\cite{b17,b41,b42}.  Although the Hoover-Leete motion equations match the Gaussian isokinetic motion equations to second order in the time the third-order motions, $\dddot q$, differ.  The Hoover-Leete approach is a somewhat different way of imposing constant kinetic energy on a Hamiltonian system.  Dettmann and Morriss derived a Hamiltonian, different to the Hoover-Leete one, which satisfies Gauss' Principle and likewise imposes a constraint on the kinetic energy.

The dynamics from the Hoover-Leete Hamiltonian is straightforward and, applied to our heat-flow problem, leads to analogs of the logarithmic-thermostat approaches.  As usual, the particle masses and Boltzmann's constant are set equal to unity.  The motion equations are less stiff, with a timestep $dt = 0.001$ conserving energy with six-digit accuracy for a billion timesteps.  Just as with the other Hamiltonian approaches to two-temperature mechanics, the cold and hot portions of the chain appear to have no influence on their Newtonian neighbors. And again, a Hamiltonian approach to thermostating fails, as shown in Figure 4.

\subsection{Configurational Thermostat}

Landau and Lifshitz' expression for configurational temperature\cite{b34},
$$
kT_c = \langle \ F^2 \ \rangle /\langle \ \nabla^2{\cal H} \ \rangle \ ,
$$
can likewise be applied to the heatflow problem.  Figure 6 of Reference 17, for a 600-particle $\phi^4$ chain, with 200 particles in each of the three regions, shows again that there is no tendency for the reservoirs to influence the temperatures of the nearby Newtonian particles.  We resist carrying out additional simulations for the model here because the equations of motion are relatively complicated, requiring, as they do, a Lagrange multiplier to control not only $\dot T \equiv 0$, but also $\ddot T\equiv 0$ .  The 600-particle results in Reference 17 lead again to the same conclusion: Hamiltonian thermostats are ineffective, unless, as is the case with the configurational temperature, the degrees of freedom being thermostated are already at the ``right'' temperature and with the first and second time derivatives of that temperature constrained to vanish.

\section{Summary and Conclusions}

Reservoirs based on Nos\'e-Hoover dynamics are an efficient and particularly successful route to nonequilibrium properties.  The underlying derivation for this dynamics can be based on Nos\'e's Hamiltonian, or on the Dettmann-Morriss Hamiltonian, or on the phase-space continuity equation necessarily obeyed by {\it any} equations of motion.  The nonHamiltonian approach provides the simplest route and has been generalized in many useful directions, even including Brownian motion\cite{b30}.  Applied to the $\phi^4$ model the Nos\'e-Hoover reservoirs generate heat flow obeying Fourier's Law, along with realistic temperature profiles.

It has been argued that heat flow is an unnecessarily demanding thermostat test\cite{b43}.  We strongly disagree.  Unless a ``thermostat'' is capable of transporting heat away from ``hot'' degrees of freedom and transferring heat toward ``cold'' ones, it is certainly unfit to ``control'' temperature.  For this reason we advocate testing any proposed thermostats with the highly-useful $\phi^4$ model investigated in the present work.

Jones and Leimkuhler\cite{b44} have recently studied the usefulness of {\it stochastic} forces in determining the usefulness (they term this ``adaptability'') of thermostats.  They seek thermostats which are both ergodic (in very small systems this is a reasonable request) and which exactly reproduce their target temperature (we agree, and view this requirement as a necessary property of any ``thermostat'' worthy of the name).  Because we believe that determinism and time-reversibility are also indispensable thermostat properties, we don't favor stochastic thermostat tests.

Here we have used the $\phi^4$ model as a test of four Hamiltonian reservoir models, [1] Nos\'e's original Hamiltonian, [2] the Logarithmic Hamiltonian reservoir of Campisi {\it et alii}, [3] the Hoover-Leete isokinetic Hamiltonian reservoir, and [4] the Travis-Braga version of Landau and Lifshitz' configurational temperature reservoir.  {\it Only the Nos\'e-Hoover reservoirs generate a realistic heat flow.}  The Hamiltonian models are relatively stiff to implement and provide no equilibration between the hot and cold reservoirs.  Evidently the twin restrictions of constant energy (Hamiltonian) and constant temperature so constrain the reservoirs that they are unable to influence even their nearest neighbors.  The failure of {\it every one} of these Hamiltonian heat flow simulations underscores the need to generalize mechanics, as did Shuichi Nos\'e, in order to treat nonequilibrium problems.  Hamiltonian mechanics itself cannot provide heat sources or sinks.  It is simply too specialized for the realistic treatment of nonequilibrium flows.

It is also noteworthy that the Hamiltonian thermostating methods require one or two orders of magnitude more computer time in order to match the accuracy of the Nos\'e-Hoover thermostat.

\section{Acknowledgment}

We thank Stefano Ruffo and Marc Mel\'endez Schofield for their encouragement as well as Arek Bra\'nka, Julius Jellinek, Paco Uribe, and Estela Blaisten-Barojas for several useful email communications and references.  We also appreciate the suggestions offered by a referee.

\end{document}